\documentclass{article}
\usepackage{arxiv}

\usepackage{times}
\usepackage{cite}
\usepackage{url}
\usepackage{subfigure}
\usepackage[pdftex]{graphicx}
\graphicspath{{Figures_CoviLearn/}}
\usepackage{xcolor}
\usepackage{rotating}
\usepackage{rotfloat}
\usepackage{amsmath}
\usepackage{amssymb}
\usepackage{algorithm}
\usepackage{algorithmic}
\usepackage{gensymb}
\usepackage{textcomp}
\usepackage{placeins}
\usepackage{balance}
\usepackage{subfigure}
\usepackage{booktabs}
\usepackage{tabularx}
\usepackage{array}
\usepackage{wrapfig}
\usepackage{picins}
\usepackage[utf8]{inputenc} % allow utf-8 input
\usepackage[T1]{fontenc}    % use 8-bit T1 fonts
\makeatother

\title{CoviLearn: A Machine Learning Integrated Smart X-Ray Device in Healthcare Cyber-Physical System for Automatic Initial Screening of COVID-19}

\author{Debanjan Das 
\\
	Dept of ECE\\
	IIIT Naya Raipur, Chhattisgarh, India \\
	\texttt{debanjan@iiitnr.edu.in} \\
\And
	Chirag Samal\\
	Dept of ECE\\
	IIIT Naya Raipur, Chhattisgarh, India \\
	\texttt{chirag18101@iiitnr.edu.in} \\
		\And
	Deewanshu Ukey\\
	Dept of CSE\\
	IIIT Naya Raipur, Chhattisgarh, India \\
	\texttt{deewanshu18100@iiitnr.edu.in} \\
	\And
	Gourav Chowdhary\\
	Dept of ECE\\
	IIIT Naya Raipur, Chhattisgarh, India \\
	\texttt{gourav18101@iiitnr.edu.in} \\
\And
	Saraju P. Mohanty\\
	Dept of CSE\\
	University of North Texas, USA \\
	\texttt{smohanty@ieee.org}	
	}
\begin{document}

\maketitle

\begin{abstract}
The pandemic of novel Coronavirus Disease 2019 (COVID-19) is widespread all over the world causing serious health problems as well as serious impact on the global economy. Reliable and fast testing of the COVID-19 has been a challenge for researchers and healthcare practitioners. In this work we present a novel machine learning (ML) integrated X-ray device in Healthcare Cyber-Physical System (H-CPS) or smart healthcare framework (called ``CoviLearn'') to allow healthcare practitioners to perform automatic initial screening of COVID-19 patients. We propose convolutional neural network (CNN) models of X-ray images integrated into an X-ray device for automatic COVID-19 detection. The proposed CoviLearn device will be useful in detecting if a person is COVID-19 positive or negative by considering the chest X-ray image of individuals. CoviLearn will be useful tool doctors to detect potential COVID-19 infections instantaneously without taking more intrusive healthcare data samples, such as saliva and blood. COVID-19 attacks the endothelium tissues that support respiratory tract, X-rays images can be used to analyze the health of a patient's lungs. As all healthcare centers have X-ray machines, it could be possible to use proposed CoviLearn X-rays to test for COVID-19 without the especial test kits. Our proposed automated analysis system CoviLearn which has 99\% accuracy will be able to save valuable time of medical professionals as the X-ray machines come with a drawback as it needed a radiology expert. 
\end{abstract}

%\begin{IEEEkeywords}
\keywords{Smart Healthcare, Healthcare-Cyber Physical System (H-CPS), Machine Learning, Deep Neural Network (DNN), Convolutional Neural Network (CNN), COVID-19, X-ray}
%\end{IEEEkeywords}

%%%%%%%%%%%%%%%%%%%%%%%%%%%%%%%%%%%%%%%
\section{Introduction}

Coronavirus disease (COVID-19) is a respiratory tract infectious disease that has spread across the world \cite{wang2020Lancet, 9174644}. Coronaviruses belong to such a family of viruses whose infection can cause complications that vary from a cold to shortness of breath. The virus contagion introduces symptoms like cough, fever, shortness of breath are its symptoms. Several patients also develop pneumonia (named novel coronavirus pneumonia, NCP) and progress rapidly towards severe acute respiratory failure with a very poor prognosis and high mortality \cite{huang2020clinical}. Global statistics suggests sizable mortality due to COVID-19. Subsequently, the pandemic nature of the coronavirus and absent of vaccine/medicine makes the diagnosis an urgent medical crisis \cite{xie2020insight}.

At present, the most common standard testing method for COVID-19 diagnosis is a real-time reverse transcription-polymerase chain reaction (rRT-PCR). In these test, a swab from patient is taken from their nose and the swab is kept in a special medium called ``virus transport medium'', to protect the RNA. Once these swabs reach the lab for testing then the swab is mixed with liquid and then these liquid is centrifuge to separate out ``pellet'' of viral cells. Now the RNA cells are separated out from these pellet and converted in two-stranded DNA using enzyme as these pellets contains very few RNA which are insufficient for testing process. After these primers along with a probe, fluorescent dye is used to bind to specific DNA sequence. Actually, the primers bind to specific DNA sequences from extremities and a probe is bound in the middle of that DNA. As both the extremities primers keep moving towards the probe, after a particular instance of time they come too near to probe and a quencher releases a fluorescent dye, signaling that the patient is positive for coronavirus \cite{jiang2020review}. The whole process takes nearly 8 hours, but results generally arrive after a day or two depending on the time taken from the swab to reach the lab.

At this point, there is a requirement of quick diagnosis, treatment and care to fight against COVID-19. In this endeavor, studies prove that COVID-19 infects the lungs which are visible in the chest X-rays and CT images, in the form of ground-glass opacities \cite{sethy2020detection}. It is observed that the affected lung from the corona COVID-19 is filled with smooth and thick mucus \cite{simpson2020radiological}. Therefore, chest X-ray is an important tool in the diagnosis of lung diseases including pneumonia. Chest X-ray is traditionally being used to determine various lung diseases such as Atelectasis, Pneumonia, Cardiomegaly, and Pneumothorax. However, the analysis of X-ray images is tedious tasks and requires expert radiologists. In this endeavor, several computer algorithms and diagnosis tools \cite{pattrapisetwong2016automatic, qin2018computer} have been proposed to get a detailed insight of the X-ray images. Recently, machine learning and deep learning based analysis has become promising tools to investigate the chest and pulmonary diseases \cite{pasa2019efficient, 8609997, drozdov2020supervised, shan2020lung}. Although the above mentioned studies have performed efficiently, however, lacks in terms of higher accuracy, computational time and error rate. Therefore, the selection of proper deep learning-based automated analyzer and predictor for coronavirus patients would be very beneficial and helpful for the medical department and society. Additionally, machine learning (ML) or deep learning (DL) approaches provide a cost effect and rapid test results as compared to insufficient number of available RT-PCR test kits.

Further, as COVID-19 is spreading more and more rapidly through person-to-person transmission, countries are facing sudden lockdown measures, forcing people to self-quarantine. All these activities have own negative effects in terms of economy, quality of life, and other health issues \cite{9085930}. Even the hospitals and healthcare professions are going through this lockdown and quarantine phase on regular basis. Clearly an alternative, remote-based, online diagnostic and testing solution is required to fill this urgent and unmet need. The Internet of Medical Things (IoMT) could be extended to achieve this healthcare-specific solution. With this motivation, the present work proposed a machine learning (ML) based Healthcare Cyber-Physical System (H-CPS) or smart healthcare framework (called CoviLearn) to allow healthcare practitioners to perform automatic initial screening of COVID-19 patients from their X-ray image data. We propose Convolutional Neural Network (CNN) models of X-ray images for automatic COVID-19 detection. These models can be used to detect if a person is COVID-19 positive or negative by considering the chest X-ray image of individuals. CoviLearn will be very useful tool for doctors to detect corona positive patients at an instant without taking samples of various other healthcare data such as saliva abd blood. COVID-19 attacks the endothelium tissues that support respiratory tract, X-rays images can be used to analyze the health of a patient's lungs. As all healthcare center have X-ray machines, it could be possible to use X-rays to test for COVID-19 without the especial test kits. Our proposed automated analysis system CoviLearn will be able to save valuable time of medical professional as the X-ray machines come with a drawback as it needed a radiology expert. Thus, CoviLearn work will be a complimentary effort in H-CPS in addition to the EasyBand and GlobeChain works which are geared towards mobility during pandemic \cite{9085930, 9416228}.

The rest of the paper is organized as follows. Section \ref{Sec:CoviLeanr_Vision} presents healthcare-CPS framework. After that, Section \ref{Sec:Prior_Works} discuss some related prior works. Subsequently, Section \ref{Sec:Novel_Contributions} details the contribution of the paper along with its novelty. Section \ref{Sec:Proposed_Solution} presents our proposed solution with details. Later, Section \ref{Sec:Experimental_Results} presents details about our experimental setup and obtained results. Finally, Section \ref{Sec:Conclusion} concludes the current study and describes some possibilities for future works.

\begin{figure*}[htbp]
\centering{\includegraphics[width=0.98\textwidth]{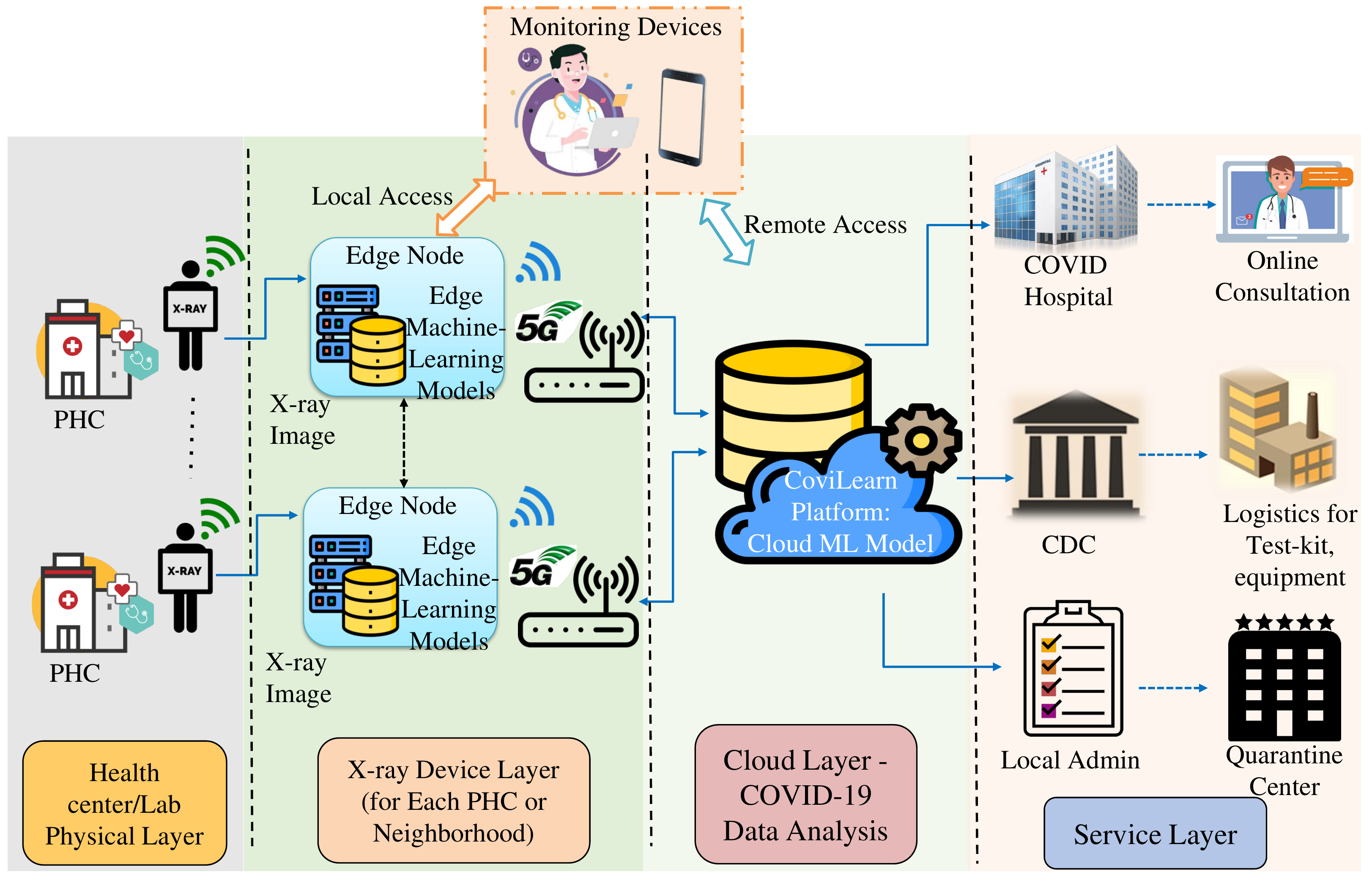}}
\caption{Schematic of the Healthcare Cyber-Physical System (H-CPS) ecosystem (CoviLearn) concept for combating COVID-19.}
\label{FIG:H-CPS_Framework_for_COVID-19}
\end{figure*}

%%%%%%%%%%%%%%%%%%%%%%%%%%%%%%%%%%%%%%%%%%
\section{Our Vision of CoviLearn in Healthcare-CPS Framework}
\label{Sec:CoviLeanr_Vision}

The proposed Healthcare-CPS (H-CPS) for CoviLearn brings the patients, doctors and test lab in a single smart healthcare platform as shown in Fig. \ref{FIG:H-CPS_Framework_for_COVID-19}. The system helps patients receive proper healthcare with minimum physical contact with others. The primary health centers/test labs could upload the test reports (chest X-ray) of patients to the IoMT platform. Radiologists are taking the help of the X-ray image to diagnose the condition of lungs and respiratory tract as a tool to start the COVID treatment.

A Deep neural network (DNN) based model, CoviLearn is implemented on the server, which could detect if a person is COVID-19 positive or negative by considering the chest X-ray image of individuals. 
These information could be transferred to nearby COVID hospitals, the Center for Disease Control (CDC), and state and local health bureaus. COVID hospitals could subsequently offer online health consultations based on the patient's condition. The givernment could allocate equipment and designate quarantine station to that patient. Therefore, using this H-CPS, people could dynamically monitor their disease status and receive proper medical needs without spreading the virus to others.

%%%%%%%%%%%%%%%%%%%%%%%%%%%%%%%%%%%%%%
\section{Related Prior Works}
\label{Sec:Prior_Works}

\subsection{Consumer Electronics for Smart Healthcare}

In the consumer electronics field, several healthcare and medical systems have been developed to monitor and diagnose the health status of a patient. Table \ref{TABLE:CE_Healthcare} summaries relevant existing research and development on healthcare-related technologies in the consumer electronics field. A perception  neural network based daily nutrition monitoring system can predict future diet \cite{sundaravadivel2018smart}. A RFID enabled personal blood pressure measurement and monitoring system has been developed for the aged person \cite{hung2012design}. iLog \cite{Rachakonda} is a personal eating behaviors management system at the interface of IoMT. This is an automatic IoT Edge level system, which provides information on the emotional state of a person along with the classification of eating behaviors to Normal-Eating or Stress-Eating. Recently, wearable devices including accelerometer, gyroscope, activity tracker, and electrocardiogram (ECG) \cite{9446547}. become very popular for monitoring and managing the personal health status such as daily activities, detection of falling event, cardiac anomaly and arrhythmia \cite{wang2016outdoor, dey2017developing}.
Similarly, research has explored to develop remote, real-time and point of care ECG monitoring at the interface of wireless sensor network and IoT \cite{raj2018personalized}.

Additionally, health data analytic is being a choice of interest in the smart healthcare system. Machine learning and deep learning based data analytics are used to extract daily behavior of office workers \cite{varma2019health}\cite{lawanot2019daily}. A new hardware friendly, convolutional neural network (CNN) model has been proposed for eye gaze estimation towards predicting the cognitive behavior of the drivers \cite{lemley2019convolutional}. Further, utilizing the advantage of data analytic (ML/DL) smartphone-based systems and Apps are being used extensively in smart healthcare applications. There are several smartphone-based applications that monitor and diagnose the lung functionality and respiration \cite{yang2020smartphone}, anemia \cite{mannino2018smartphone}, pressure \cite{aloudat2019automated}, the skin lesions \cite{chao2017smartphone}, as well as heart functionality test \cite{faezipour2010patient}, cognitive analysis \cite{aljaaf2019patients}, and several more \cite{vinay2013smartphone}. These studies provide customized healthcare services over IoT network for individual user based on either smartphones or smartwatch collected data analysis.

\begin{table*}[htbp]
\caption{A summary of Consumer Electronics on Smart Healthcare}
\label{TABLE:CE_Healthcare}
\centering
%\begin{tabularx}{\linewidth}{|X|X|X|}
%\begin{tabular}{|p{3.8cm}|p{5cm}|p{4cm}|}
\begin{tabular}{p{3.4cm} p{6.1cm} p{4.5cm}}
\hline
%\hline
\textbf{Works} & \textbf{Main Features} & \textbf{Remarks} \\
\hline
\hline 
%hline
%Sundaravadivel et al. \cite{sundaravadivel2018smart} & Bayesian network to extract nutrient features for detecting nutritional balance of user & IoMT-based automated 
%\\
%\\
%\hline
Hung et al. \cite{hung2012design} & Blood pressure measurement with a health management system & Applicable for remote healthcare center 
\\
\\
%\hline
Wang et al. \cite{wang2016outdoor} & Healthcare box (H-Box) enabling outdoor ECG monitoring with automatic medical attention & Intelligent, interactive and features automatic learning system for elderly care
\\
\\
%\hline
Dey et al. \cite{dey2017developing} & Zigbee wireless sensor networked ECG monitoring & Smart home healthcare 
\\
\\
%\hline
Varma et al. \cite{varma2019health} & IMU sensor based healthband for identification different bad habits of user & Health assessment of office workers
\\
\\
%\hline
Rachakonda et al. \cite{Rachakonda} & Wearable iLog-Glasses automatically detect, classify and quantify the food using an edge computed deep learning model & Automatic IoMT Edge level system to help a person differentiate between Stress-Eating and Normal-Eating
\\
\\
%\hline
Sayeed et al. \cite{sayeed2019eseiz} & Signal rejection algorithm (SRA) and IoMT & Low power remote EEG monitoring and detection of seizure
\\
\\
%\hline
\textbf{Proposed System (CoviLearn)} & Smart X-ray imaging and H-CPS & X-ray based automatic initial screening of COVID-19 
\\
\hline
%\end{tabularx}
\end{tabular}
\end{table*}

\subsection{X-Ray based Solutions Smart Healthcare}

Recent advancements of AI, machine learning (ML) and deep learning (DL) in smart healthcare applications \cite{esteva2019guide, topol2019high, ahad20195g} have inspired innovations in the development of novel AI-based radiological diagnostic technology. Therefore, at the start of the pandemic in early January, reports from Wuhan indicated that the chest X-ray played a vital role in determining SARS-CoV-2 virus commonly known as Corona Virus \cite{ai2020correlation}. Subsequently, RT-PCR (Real Time Polymerase Chain Reaction) test is done to confirm the presence of coronavirus antibody \cite{jiang2020review}. Chest X-ray is traditionally being used to determine various lung diseases such as Atelectasis, Pneumonia, Cardiomegaly, and Pneumothorax. However, the most significant challenge with chest X-ray is human dependent detection of these X-ray. The trained radiologist needs to have expertise to differentiate between the heterogeneous colour distribution of air, while that flows through the lungs. To overcome this a machine learning based system was developed which uses heat maps technique over the X-ray images denoting various diseases \cite{bruno2019using}.

\begin{table*}[htbp]
\caption{Comparative Perspective with Related AI Works for COVID-19 Detection}
\label{TABLE:Lit}
\centering
%\begin{tabularx}{\linewidth}{|X|X|X|}
%\begin{tabular}{|p{3.8cm}|p{5cm}|p{4cm}|}
\begin{tabular}{p{4.1cm} p{5.5cm} p{4cm}}
\hline
%\hline
\textbf{Technology} & \textbf{Key contributions} & \textbf{Remarks} 
\\
\hline
\hline
Lung Infection Quantification using CT Images, 2020 \cite{shan2020lung} & Deep learning model for differentiating heterogeneous colour distribution of air & Human dependent detection  
\\
\\
%\hline
Deep learning to screen COVID-pneumonia, 2020 \cite{12_butt2020deep} & 3D -DL model for pulmonary CT scan Location-attention classification model & Accuracy limited to 86.7\% 
\\
\\
%\hline
CNNs for chest X-Ray tuberculosis screening, 2019 \cite{3_pasa2019efficient} &  A deep learning architecture tailored to tuberculosis diagnosis & Study limits to tuberculosis chest X-ray
\\
\\
%\hline
COVID-Net, 2020 \cite{wang2020covid} & Presented an open source network COVID-Net, which shows more accurate that VGG-19 and ResNet-50 for COVID-19 detection & Accuracy limited to 93.3\%
\\
\\
%\hline
Detection of pneumonia in chest X-ray images, 2011 \cite{parveen2011detection} & detection of pneumonia by unsupervised fuzzy c-means classification learning algorithm & Color based marking due to pneumonia in X-ray image
\\
\\
%\hline
DarkNet and YOLO, 2020 \cite{ozturk2020automated} & NCP detection using 500 pneumonia images and 500 non-pneumonia images & Classification accuracy of 98.08\% for binary classes and 87.02\% for multi-class cases
\\
\\
%\hline
\textbf{CoviLearn (Current Paper)} & CNN based H-CPS classifies COVID-19 and Normal & Accurate with potential IoMT based H-CPS solution  
\\
\hline
%\end{tabularx}
\end{tabular}
\end{table*}

There are several ML/DL-based models have been presented in existing literature for diagnosis using X-ray and CT images. Table \ref{TABLE:Lit} presents a summary of the studies described in this section, focusing on their most important characteristics. 
%Chen et al. \cite{chen2020quantitative} has reviewed various quantitative models of chest CT images and showed effectiveness of these tools in accurate diagnosis. 
Few groups have already extended these advanced ML/DL algorithms for the diagnosis of COVID-19 \cite{chen2020quantitative}. CT images have been analyzed for COVID-19 infected patients using a deep learning model with an accuracy of 79.3\% . A Convolutional Neural Network (CNN) based COVID-Net architecture has been presented for detection of disease with a test accuracy of 92.4\% \cite{wang2020covid}. Pulmonary CT scan images to make a 3-D deep learning model has been explored to solve the same problem \cite{butt2020deep}. 
%It uses a location-attention classification model which uses noisy- Bayesian function  to identify corresponding confidence scores and infection type. 
%Deep learning has been explored to identify drug suitable for coronavirus \cite{wang2020abnormal}. 
%Therefore, they use the model with the RNA sequence of coronavirus to predict which drug is best for coronavirus treatment and come up with suggestion that Adenosine, Vidabrine might help. 
One of the major challenges of these techniques to achieve higher accuracy. But, since the chest X-ray dataset of coronavirus is very limited there is one common problem of class imbalance while training the model i.e. less number of coronavirus images as compared to normal lung images, in all the related work on coronavirus detection. This problem of dataset imbalance results in less accuracy of the model and thus becomes less efficient.

%%%%%%%%%%%%%%%%%%%%%%%%%%%%%%%%%%%%%%%%%%%%%%%
\section{Contributions of the Current Paper}
\label{Sec:Novel_Contributions}

We present a AI based Healthcare Cyber-Physical System (H-CPS) (called CoviLearn) to allow healthcare practitioners to perform automatic initial screening of COVID-19 patients from their X-ray image data. The paper specifically aims for identifying the best learning model available for covid detection using chest X-ray images by comparing different deep learning models on the basis of different parameters like accuracy, sensitivity, specificity and deploying it for fast, accurate and hassle-free covid detection by incorporating health-CPS.

\subsection{Problem Addressed in the Current Paper}

As with the current nasal swab testing method it takes around 8 hours for the test to show results, and generally takes a day for the test results to come. Therefore there should be a fast alternative for nasal swab COVID-19 testing which is accurate enough for identifying COVID-19. One such alternative is chest X-ray images, but to draw inference from chest X-ray requires trained radiologists. The present work presented a human independent COVID-19 detector by applying deep learning on chest X-ray images and predict results within seconds.

\subsection{Solution Proposed in the Current Paper}

In the present work, we have explored the deep learning technique for investigation and diagnosis of X-ray images leading to COVID-19. We have used Deep Neural Network (DNN) models to detect COVID-19. The performance parameters of these models have been characterized and found it is more reliable, accurate and specific. The machine learning trained model will provide a solution by analyzing the X-ray images of the patients, easy to function, convenient to operate. By just uploading X-ray images, the model will automatically identify the symptoms and will show perfect unbiased results.

\subsection{Novelty of the Proposed Solution}

\begin{itemize}
\item 
An architecture of Health-CPS framework to combat COVID-19 diagnosis.

\item Proposed a next generation smart X-ray machine architecture at the interface of Health-CPS.

\item The proposed Deep neural network model is an automatic tool with an end-to-end structure without manual extraction of the features from Chest X-ray images.

\item The proposed model is an efficient Heuristic search approach which can automatically find with an optimal feature-subset present in Chest X-ray images.

\item We state that the DNN with DenseNet121 convolution neural network blocks is more reliable, accurate, and specific pre-trained model as compared to the other three models in the detection of COVID-19 from Chest X-ray images.

\item Proposes an efficient deep learning network which is optimal, rational and accurate and can be easily integrated into embedded and mobile devices. This could help the health practitioners for diagnosis of the COVID-19.
\end{itemize}

\begin{figure*}[htbp]
	\centering{\includegraphics[width=0.99\textwidth]{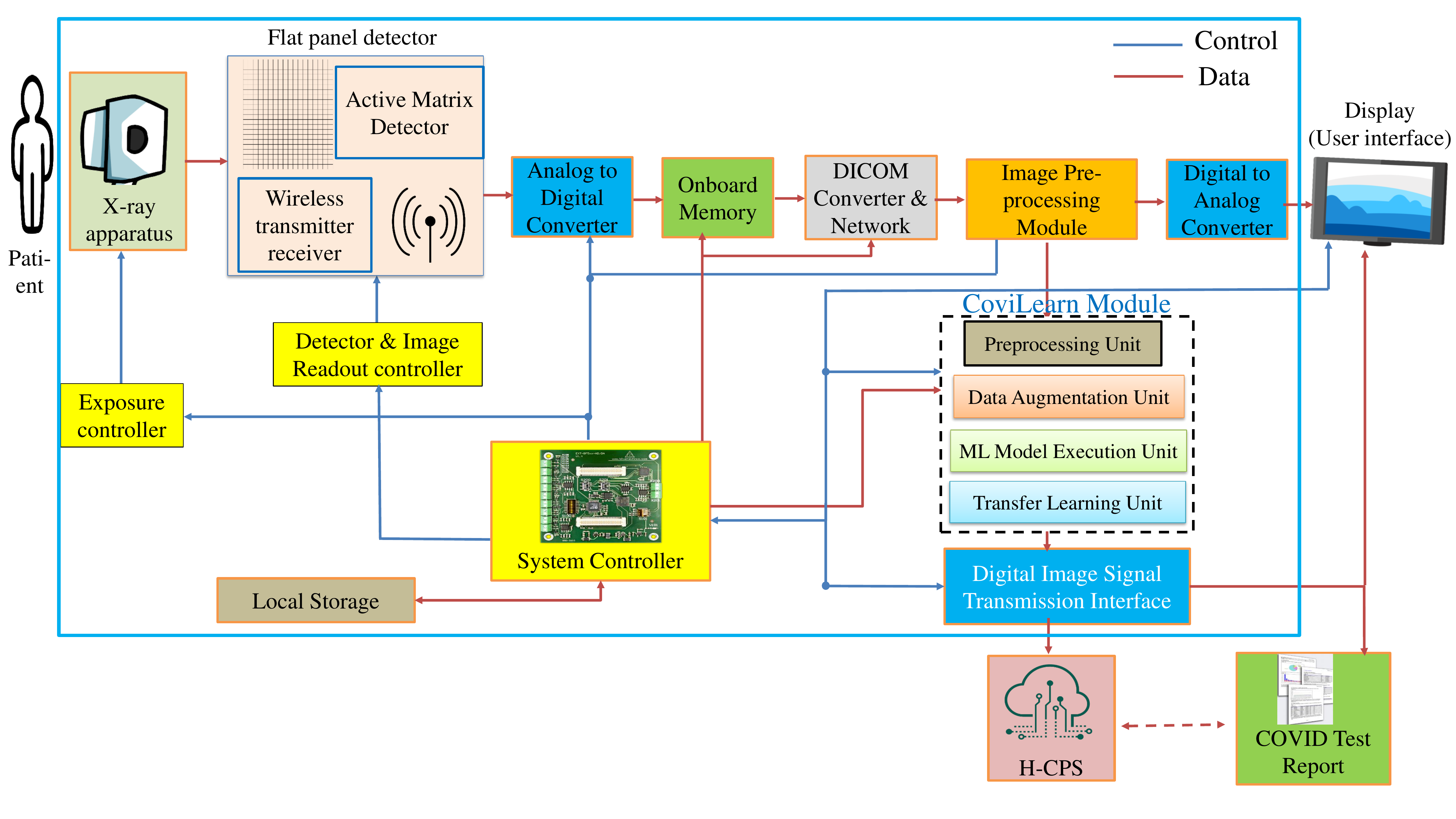}}
	\caption{The proposed next generation X-ray device (CoviLearn) integrated with machine learning models.}
	\label{FIG:NextGenX-ray}
\end{figure*}

%%%%%%%%%%%%%%%%%%%%%%%%%%%%%%%%%%%%%%%%%%%%%%%
\section{Proposed CoviLearn Device for Next Generation X-ray}
\label{Sec:Proposed_Solution}

Due to the pandemic nature of infectious diseases like COVID-19, traditional healthcare systems are not able to provide the necessary services to everyone. COVID and other pneumonia diseases can be initially screened and diagnosed by chest X-ray diagnosis. However, the existing diagnostic is suffered from limited access and a lack of radiologists. This motivates for developing next-generation X-ray devices in a Healthcare-cyber physical system perspective. The H-CPS and IoMT together will bring all the necessary agents of smart healthcare in a universal communication and connectivity platform. This H-CPS linked technologies can extend the efficiency services, such as telemedicine and multimodal simultaneous medical care.

Fig. \ref{FIG:NextGenX-ray} shows the system-level block diagram of the next generation X-ray machine integrated with CoviLearn for automatic screening of infectious diseases. It identifies most of its components, such as X-ray apparatus (tube), flat panel detector, onboard memory, DICOM protocol converter, Image processing, CoviLearn diagnosis, wired/wireless data communication, display or user interface, along with system controller. In the proposed X-ray machine, X-ray image is captured by an array of sensor in the Digital and Radiography flat panel detector. The flat panel also includes the devices of communication to next stages. The image is then saved and converted to DICOM X-ray image. Subsequently, the image is processed and based on the quality and requirement the exposure of the X-ray tube is adjusted. The captured image is stored temporarily in the local memory, after which it is
displayed on monitor screen with the help of the controller. After acquiring the quality assured image, it is then transferred to the CoviLearn model which automatically classify the image either as normal or COVID-19 affected. The image classification is performed either locally in presence of sufficient resources or on cloud by transmitting the images over network. This test results automatically sync with the H-CPS platform for necessary medical and administrative actions. The controller unit is responsible for controlling the entire sequence of events.

%%%%%%%%%%%%%%%%%%%%%%%%%%%%%%%%%%%%%%%%%%%%%%%
\section{Proposed Machine Learning based Models for CoviLearn}

\begin{figure}[htbp]
\centering{\includegraphics[width=0.60\textwidth]{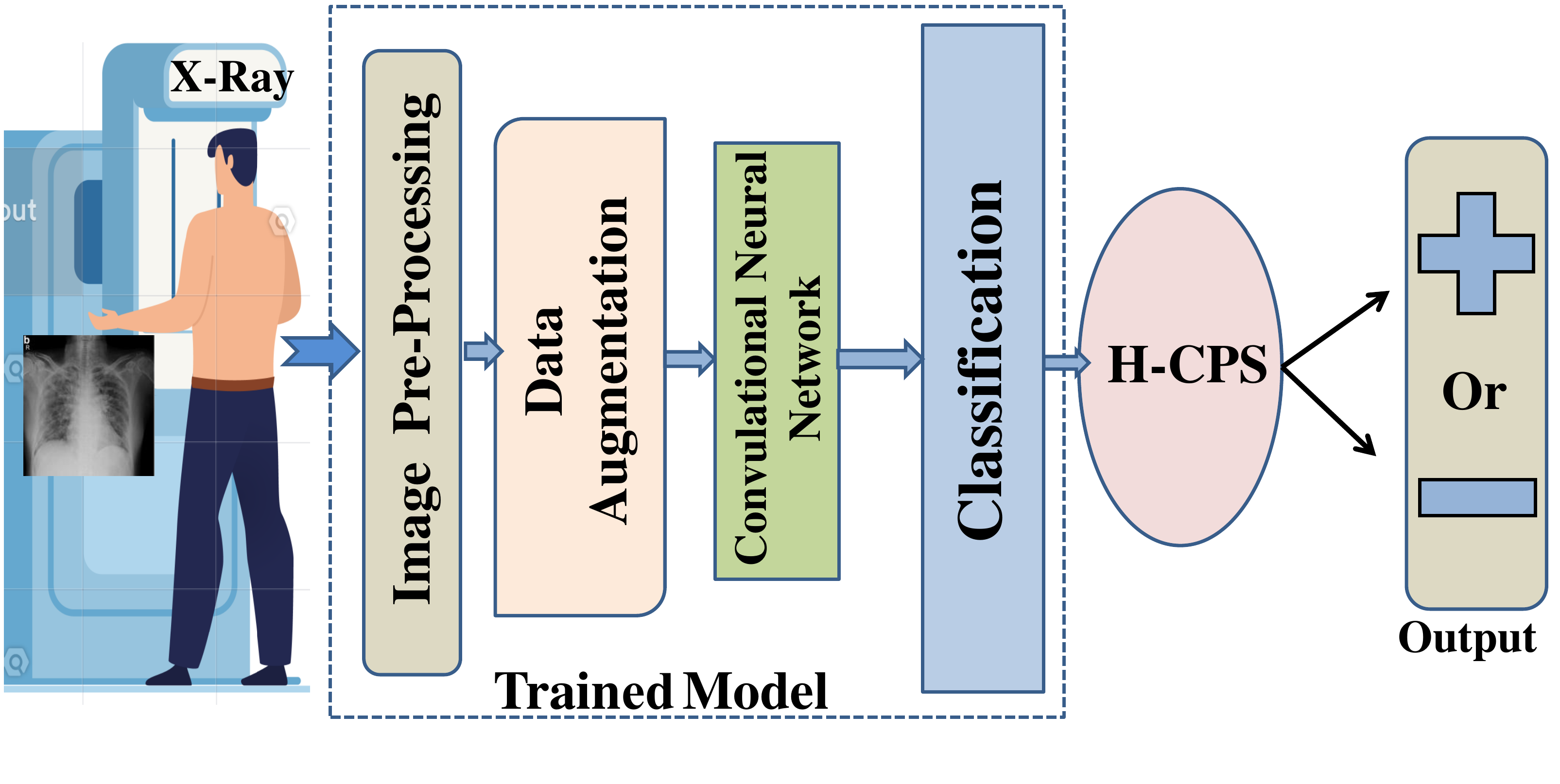}}
\caption{Process flow chart of CoviLearn for classification.}
\label{FIG:CoviLearn_Classification_Process_Flow}
\end{figure}

Fig. \ref{FIG:CoviLearn_Classification_Process_Flow} shows the process of proposed CoviLearn system for classification of COVID-19 from chest X-ray images. The detail of individual steps is discussed in subsequent subsections.

\subsection{Pre-Processing}

Since all the images are of different sizes so we have to do data pre-processing. The pre-processing is done in 3 steps i.e. The individual data is normalized by subtracting the mean RGB values, then all the pixels in an input data is converted in the range 0 to 1. Now the input data or tensor is reshaped into the appropriate tensor so that it could fit in the model, in this case the tensor is reshaped into $224 \times 224$ pixels.

\subsection{Data Augmentation} 

Deep learning models are ravenous for data and since our model only has around 250 images for each class hence the volume of our data needs to be increased and this can be achieved through data augmentation. Therefore, we augmented our images by random crop, adjust contrast,  flip, rotation, adjust brightness, horizontal and vertical shift and flip, aspect ratio, random shear, zoom and pixel jitter \cite{shorten2019survey}. Thus, it makes our system more efficient.

\subsection{The Proposed Transfer Learning for Deep Neural Network (DNN) Models} 

\begin{figure}[htbp]
\centering{\includegraphics[width=0.85\textwidth]{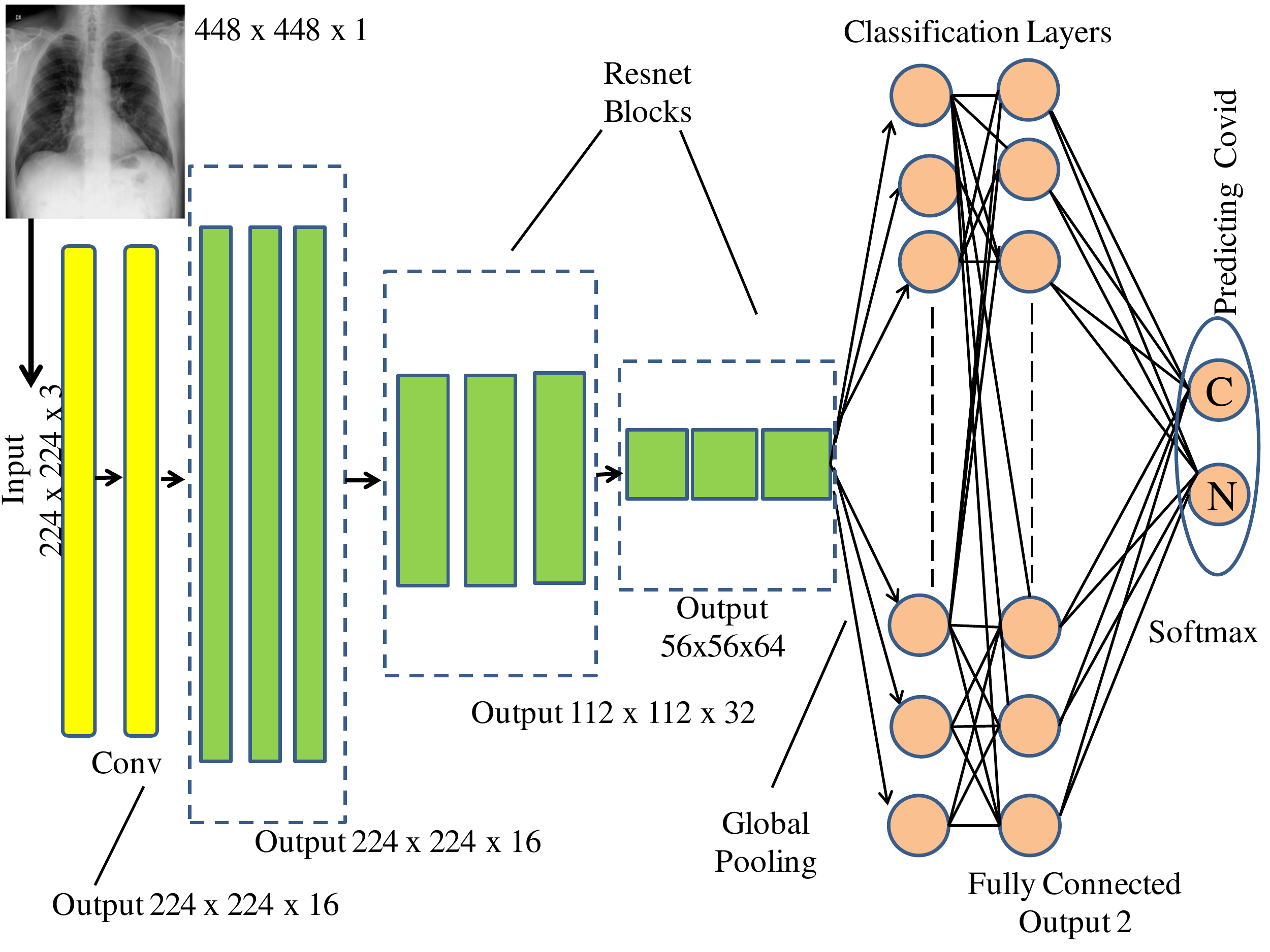}}
\caption{Organization of the DNN with classification layers.}
\label{FIG:Res}
\end{figure}

The proposed CoviLearn uses transfer learning to predict the results i.e. the patient is coronavirus positive or not. Transfer learning removes the requirement of large dataset and has been used in different applications, such as healthcare, and manufacturing. Transfer learning uses the knowledge learned in training the large dataset and transfers that knowledge in the smaller dataset. In the present work, we have used four different DNN models with different blocks to train our network and compared the results, so that we can use the model with highest accuracy. The DNN models uses some of basic blocks of ResNet50, ResNet101, DenseNet121 and DenseNet169 models. Detailed structural organizational of network layers is shown in Fig. \ref{FIG:Res} where DNN model is divided into phases, starting from taking input as a image and then training the model by sequentially passing the set of images into convolutional networks with the help of algorithms followed by a classification layer to predict utmost accurate results.

\subsection{Proposed Flow for Training and Testing}

Overall, the CoviLearn model at first reads the input image then swap the color channels and resize it to $224 \times 224$ pixels. Afterward, the data and label list are converted to NumPy array and the pixel intensity is set between 0 to 1 by dividing the input image by 255. Subsequently one-hot encoding is performed to the labels. Thereafter, various models are loaded one at a time by freezing few upper layer and a base layer is created with dropout. Then the input tensor of size $ 224 \times 224$ is loaded into the model and compiled using Adam optimizer and binary cross entropy loss. The detail process is described in Algorithm \ref{Algo 1}.

\begin{algorithm}[htbp]
\caption{Training and prediction using CoviLearn.}
\label{Algo 1}
\begin{algorithmic} 
\REQUIRE ${a: \textrm{data[0]}}, {b: \textrm {label[1]}}, {z: \textrm{number of images}},$\\
${m,n: \textrm{image dimension}},{f: \textrm{basemodel}}, {g: \textrm{headmodel}}$ \\
%\STATE $img \leftarrow Capture Image$
%\STATE $m,n \leftarrow Dimension(img)$
%\STATE $gray_{img} \leftarrow Grayscale (img)$
%\STATE $blur_{img} \leftarrow Blur (gray_{img})$
%\STATE $th_{img} \leftarrow Threshold (blur_{img}, threshold  constant=90)$
%\STATE $Contours \leftarrow Find contours$

\FOR {$i=0 \ \TO \  z-1$}
\STATE $b \leftarrow image_{labelvalue}$
\STATE $image \leftarrow image_{convertcolour}$
\STATE $image \leftarrow image_{resize}(m,n)$
\STATE $a \leftarrow image$
\STATE $i++$
\ENDFOR

Model(a):
\STATE $\quad \quad f \leftarrow ResNet50, 101; DenseNet121, 169$
    \STATE $ \quad \quad g\leftarrow f.out$
    \STATE $ \quad \quad g\leftarrow Conv(g)$
    \STATE $ \quad \quad g\leftarrow Pooling(g)$
    \STATE $ \quad \quad g\leftarrow Flatten(g)$
    \STATE $ \quad \quad g\leftarrow Dense(g)$
    \STATE  $ \quad\quad\textbf{return} \ metrics_{g}$
    
Predict(test data):
	\STATE $ \quad \quad Result\leftarrow Model.predict$
    \STATE  $ \quad\quad\textbf{return} \ Result$
\end{algorithmic}
\end{algorithm}

%%%%%%%%%%%%%%%%%%%%%%%%%%%%%%%%%%%%%%%%%%%%
\section{Experimental Results}
\label{Sec:Experimental_Results}

\subsection{Experimental Setup} 

To compare the performance of different models and to choose the best out of it, three evaluation parameters  namely accuracy, sensitivity and specificity have been taken into consideration. As the test images are converted into $224 \time 224$ tensor the model predicts the above mentioned three metrics. A table is shown comparing the results between 4 models i.e. DNN I, DNN II, DNN III, DNN IV. A confusion matrix is also drawn comparing True Positive, True Negative, False Positive and False Negative values. Moreover, 4 loss/accuracy versus epoch graph is shown which illustrates how the training loss, validation loss, training accuracy and validation accuracy vary with the epoch count.

\subsection{Datasets} 

To overcome the problem of class imbalance we have manually collected 240 chest X-ray of patients having coronavirus, these images are from various resources i.e. pyimagesearch, radiopedia, sirm and eurorad. For the normal chest X-ray we have used  chest x-ray dataset from National Institute of Health (NIH), USA  \cite{ChestXray-NIHCC_URL} and filtered the normal chest X-ray from that dataset which sums up to 250 normal chest X-ray images. Subsequently, the dataset has been splited into two classes as we are doing binary classification i.e. the patient is having coronavirus or not. The images are split in 80:20 ratio for training and testing, respectively.

\subsection{Result Analysis} 

In context of coronavirus detection, True Positive (TP) is when the patient has coronavirus and the model detects coronavirus, True Negative is when the patient doesn't have coronavirus and the model also predicts the same, False Positive (FP) is when the the patient doesn't has coronavirus but the model predicts so and False Negative (FN) is when the patient has coronavirus but the model shows contrary result. Accuracy specifies the proportion of corona patients to that of non-corona patients. The metrics are expressed as follows \cite{Rachakonda}:
\begin{eqnarray}
%\begin{array}{l}
Accuracy & = & \frac{{TP + TN}}{{TN + TP + FP + FN}}\\
Sensitivity & = &  \frac{{TP}}{{TP + FN}}\\
Specificity & = &  \frac{{TN}}{{TN + FP}}
%\end{array}
\end{eqnarray}

Further, sensitivity of test is the ability to identify coronavirus patients correctly and the specificity of test is the ability to identify non-coronavirus patients correctly. Table \ref{table:comparison} summarizes the performance matrix for different deep learning model tested for the different classification schemes. DNN III, which has DenseNet121 architecture performs best over other models in classification yielding an accuracy of 98.98\%, sensitivity of 100\%, and specificity value of 98\%. Whereas, DNN I shows lowest performance value with an accuracy of 95.92\%, sensitivity of 95.83\%, and specificity value of 96\%.

\begin{table*}[htbp]
 \centering
%    \captionsetup{justification=centering}
   \caption{Performance metrics for different deep learning techniques.}
 \label{table:comparison}
%\small\addtolength{\tabcolsep}{-2pt}
%\begin{tabular}{|c|c|c|c|c|c|}
	\begin{tabular}{c c c c r c}
	\hline
\hline
\textbf{Models Explored} & \textbf{Accuracy} & \textbf{Sensitivity} & \textbf{Specificity} & \textbf{Total Parameter} & \textbf{AUC Area} \\[0.5ex] 
\hline
%\hline 
DNN I & 0.9592 & 0.9583 & 0.9600 & 23,696,066 & 0.959 
\\
\\
%\hline
DNN II & 0.9694 & 0.9792 & 0.9600 & 42,757,826 & 0.970
\\
\\
%\hline
DNN III & 0.9898 & 1.0000 & 0.9800 & 7,103,234 & 0.990
\\
\\
%\hline
DNN IV & 0.9796 & 1.0000 & 0.9600 & 12,749,570 & 0.980
\\
\hline
\end{tabular}
\end{table*}

\begin{figure}[htbp]
\centering{\includegraphics[width=0.65\textwidth]{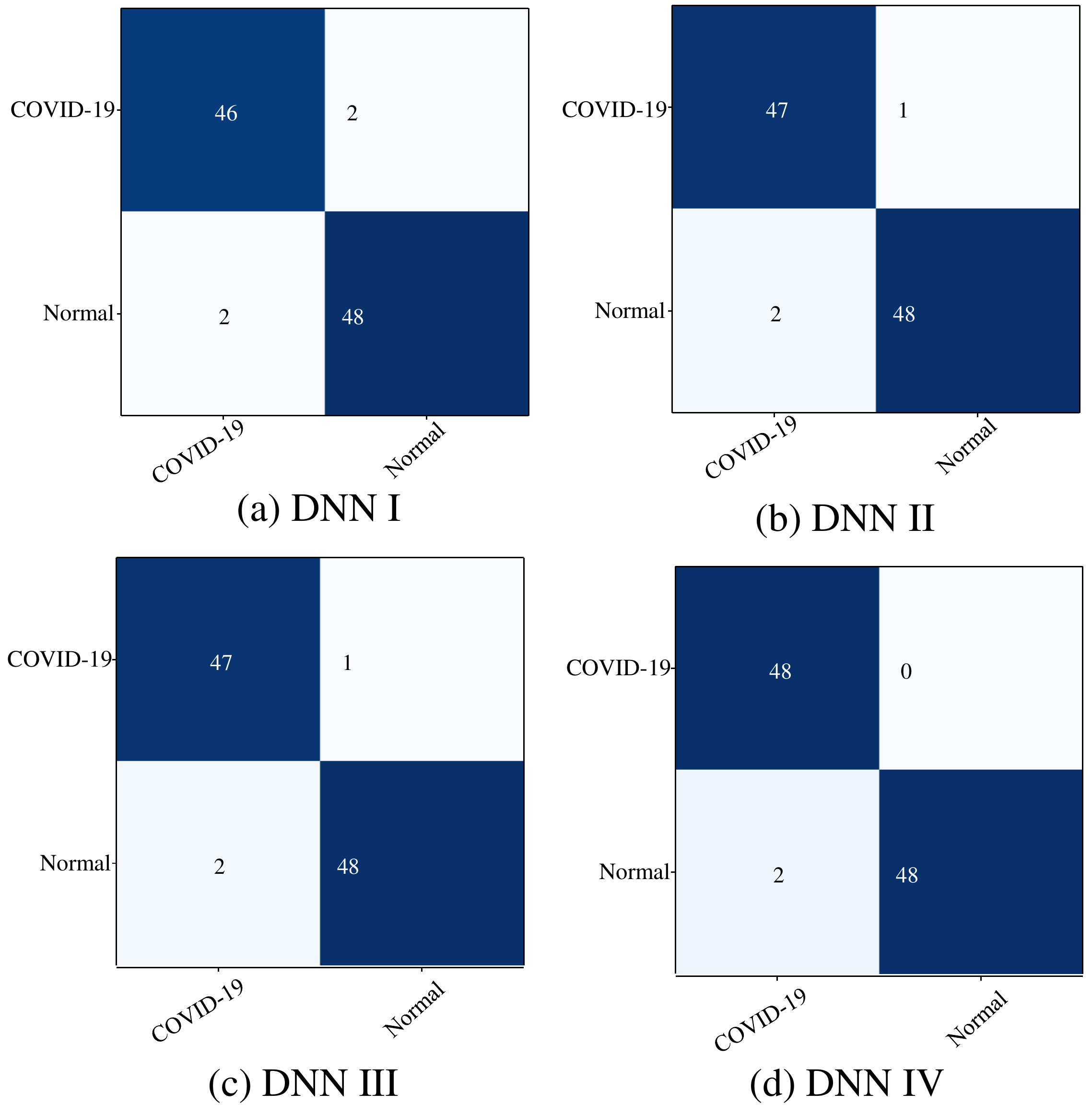}}
\caption{Confusion matrix for (a) DNN I, (b) DNN II, (c) DNN III and (d) DNN IV.}
\label{FIG:Confusion_Matrix_Results}
\end{figure}

\begin{figure}[htbp]
\centering{\includegraphics[width=0.65\textwidth]{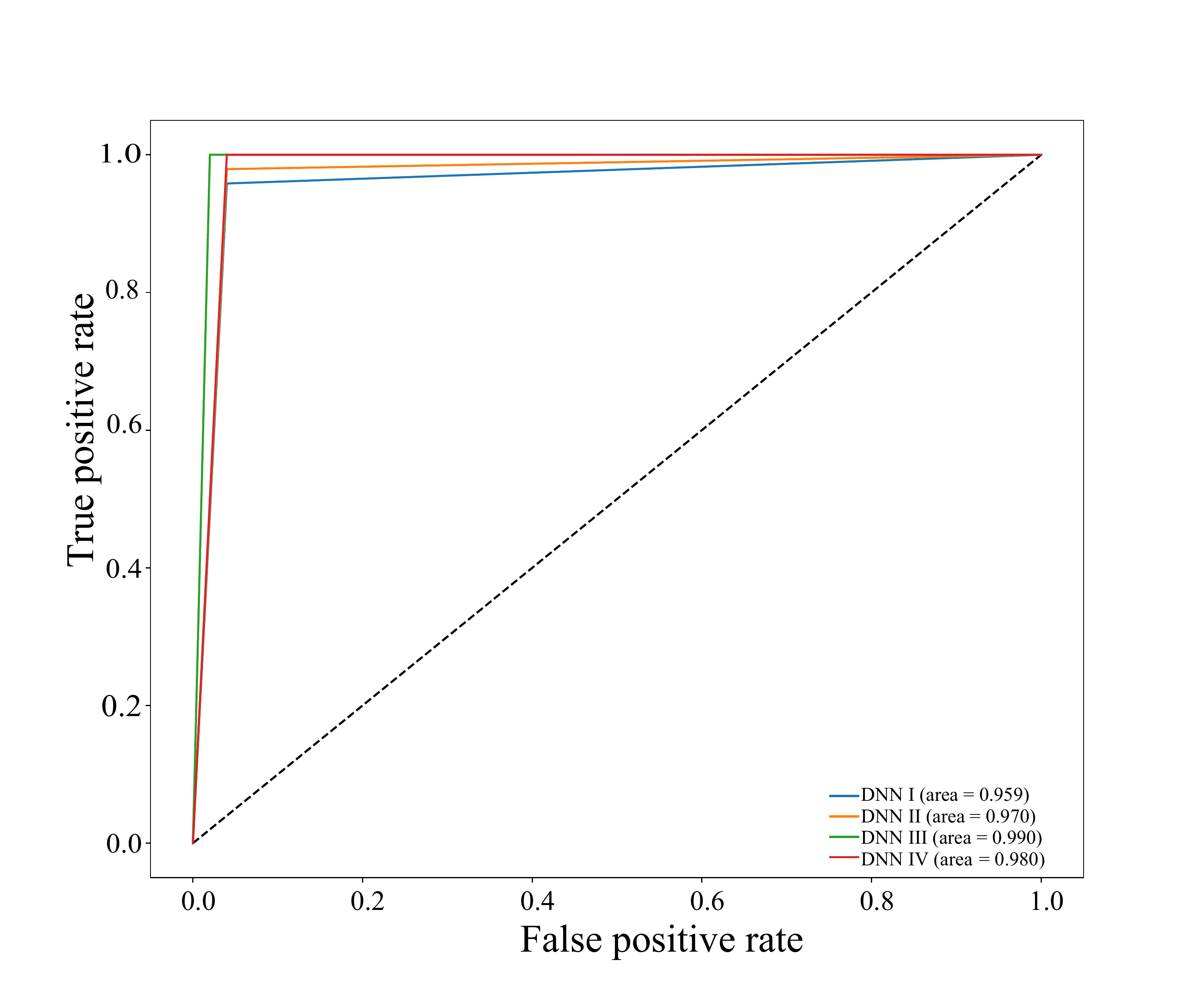}}
\caption{Comparison of the receiver-operating characteristics (ROC).}
\label{FIG:ROC_Plot}
\end{figure}

Fig. \ref{FIG:Confusion_Matrix_Results} shows the confusion matrices of COVID-19 and normal test results of the different pre-trained models. The graphs show a well defined pattern as the training and validation accuracy increases and the training and validation loss decreases as the epochs increase. Keeping into the consideration of limited computational resources, the comparison between different parameters is done for 25 epochs only. Besides the confusion matrix, receiver operating characteristic curve (ROC) plots and areas for each model is given in Fig. \ref{FIG:ROC_Plot}. DNNs which are trained with DenseNet pre-trained blocks appear to be very higher than DNN trained with ResNet blocks, with DNN III is the highest AUC area of 99\%. One of the interesting findings is the DNN which used with the ability of the DenseNet model achieves higher sensitivity and specificity. This will ensure in reduction of false positive cases for both COVID-19 and normal healthy classes.

\begin{figure}[htbp]
\centering{\includegraphics[width=0.65\textwidth]{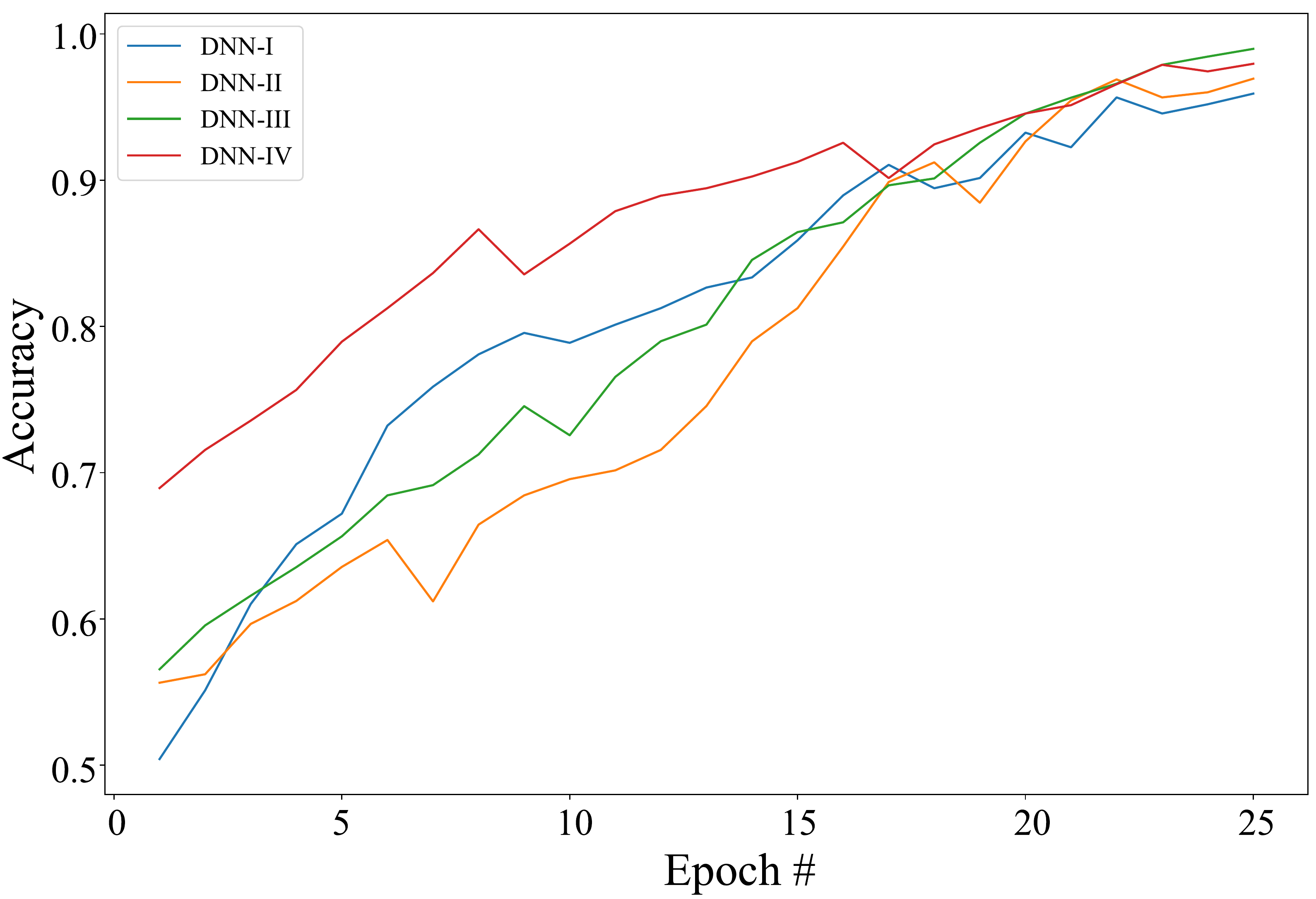}}
\caption{Classification accuracy in the deep learning system validation.}
\label{FIG:accuracy}
\end{figure}

\begin{figure}[htbp]
\centering{\includegraphics[width=0.65\textwidth]{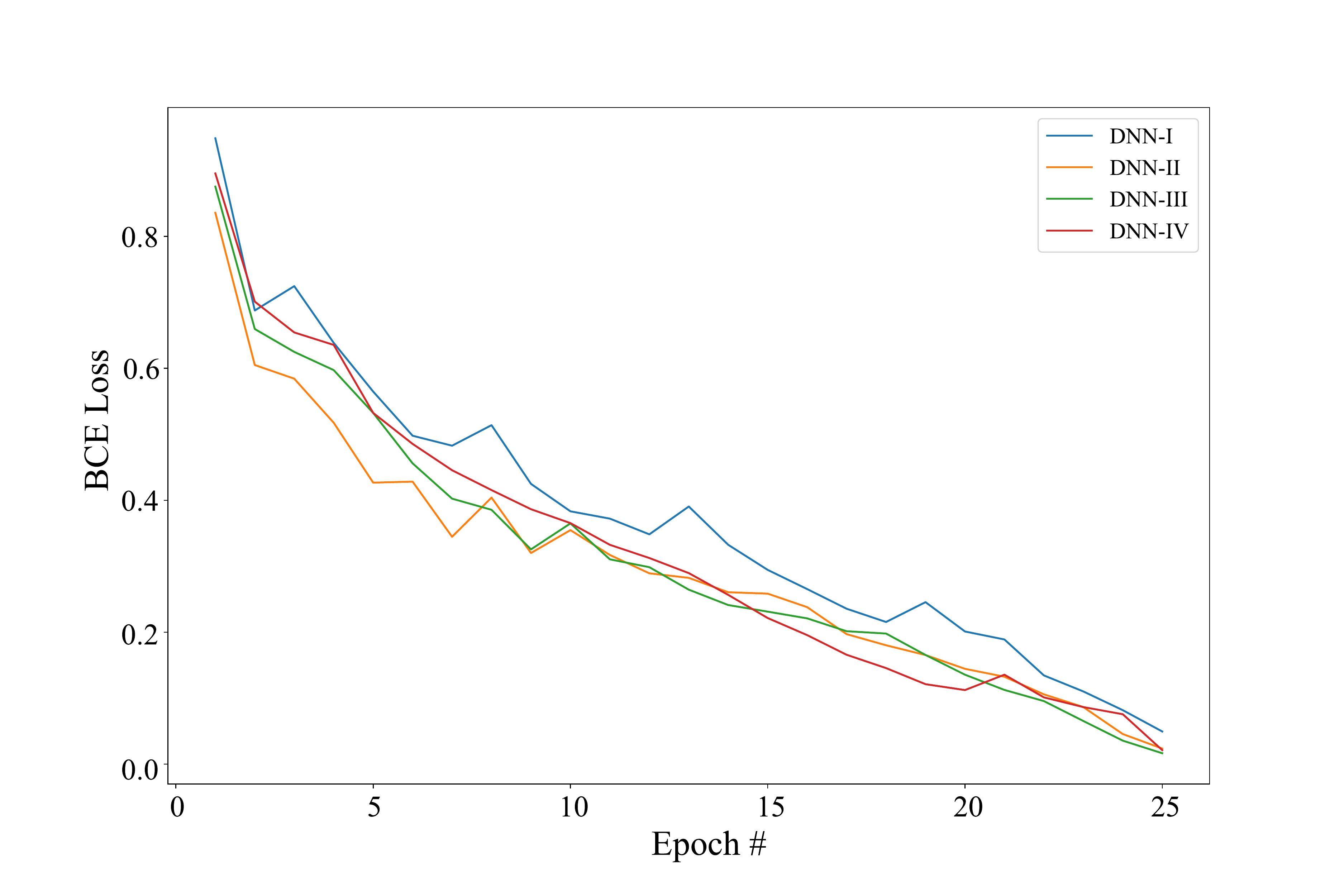}}
\caption{Binary cross entropy loss in the deep learning system validation.}
\label{FIG:loss}
\end{figure}

As evident from the relationship between accuracy and epoch, DNN-III shows the highest accuracy followed by DNN-IV, DNN-II and DNN-1. The accuracy increases with each subsequent epoch except at few epochs shown in Fig. \ref{FIG:accuracy}. A similar trend is shown in loss graphs where the loss decreases with each subsequent epochs and a similar trend is followed i.e. DNN-III shows the lowest loss followed by DNN-IV, DNN-II and DNN-1 shown in Fig. \ref{FIG:loss}.

Further, the results of the present work is compared with the few of already published literature as shown in Table \ref{tab:restComp}. In \cite{butt2020deep}, te detection of COVID-19 using classification of CT samples by CNN models with a an accuracy of 86.7\%, sensitivity of 98.2\% and a specificity of 92.2\%. CovidNet work in \cite{wang2020covid} has an accuracy of 93.3\%. The CNN based DarkCovidNet model to detect COVID-19 from chest X-ray has an accuracy of 98.08\% \cite{ozturk2020automated}. The comparison of others works in this regards is summarized in Table \ref{tab:restComp}.

\begin{table}[htbp]
\centering
\caption{Comparison of results with other recent similar works}
\label{tab:restComp}
%\begin{tabularx}{\linewidth}{|X|X|X|}
%\begin{tabular}{|p{2cm}|p{3cm}|p{2cm}|}
\begin{tabular}{p{4.2cm} p{6.0cm} p{2.2cm}}
\hline
%\hline
\textbf{Works} & \textbf{Technique} & \textbf{Accuracy (\%)} \\
\hline
\hline
Butt at al. \cite{butt2020deep} & Deep CNN model 3D -DL model & 86.7 
\\
\\
%\hline
Wang et al. \cite{wang2020covid} & CovidNet, VGG-19 and ResNet-50 model & 93.3
\\
\\
%\hline
Ozturk et al. \cite{ozturk2020automated} & DarkNet and YOLO & 98.08
\\
\\
%\hline
Khatri et al. \cite{khatri2020pneumonia} & EMD approach & 83.30
\\
\\
%\hline
Togacar et al. \cite{tougaccar2019deep} & Deep CNN model & 96.84
\\
\\
%\hline
\textbf{CoviLearn (Current Paper)} & Deep-CNN based DensNet & 98.98\\
\hline
%\end{tabularx}
\end{tabular}
\end{table}

%%%%%%%%%%%%%%%%%%%%%%%%%%%%%%%%%%%%%%%%%%%%%
\section{Conclusions and Future Scope}
\label{Sec:Conclusion}

The study presents a DNN based transfer learning approach in Healthcare Cyber-Physical System (H-CPS) framework, CoviLearn to perform automatic initial screening of COVID-19 patients from their X-ray image data. Further, it presented an architecture of next generation smart X-ray machine for automatic screening of COVID-19 like diseases at the interface of H-CPS. Four different DNNs with ResNet50, ResNet101, DenseNet121 and DenseNet169 blocks were trained and tested for classification of chest X-ray images from healthy and corona diseases infected patients. DNN III shows the highest accuracy close to 98.98 \% followed by DNN IV, DNN I and DNN II. Similarly, the sensitivity of DNN III and DNN IV is 100 \% which shows there ability to classify the deadly coronavirus correctly and that of DNN I and DNN II are close to 97\% and the highest specificity of DNN III is 98\%. Therefore, by looking at the overall robustness of DNN III, we can conclude that it is the best model of the detection of coronavirus.

The present CoviLearn platform will be very useful tool doctors to diagnosis the coronovirus disease at a low-cost, rapid, and automatically. However, further study and medical trial is required to full proof the extracted features extracted by machine learning as reliable biomarkers for COVID-19. Further, these machine learning models can be extended to diagnose other chest-related diseases including tuberculosis and pneumonia. A limitation of the study is the use of a limited number of COVID-19 X-ray images. Therefore, the future works include to create a much larger dataset and building a cloud based system where this model could be used by doctors as well as common people. In fact X-ray image solely could not confirm the COVID-19 disease, but it can be used to detect highly prone corona positive patients in the timely application of quarantine measure, until RT-PCR test examination performed.  Further, CoviLearn needs to be integrated tightly with contact tracing as well as blockchain-based privacy aware solutions in H-CPS to most effective to control the spread of pandemic while facilitating mobility \cite{9085930, 9416228, 9130396, 9353683}.

%\textbf{\textcolor{red}{Include Future Directions.}}

%\ack{The authors thank Professor James McLoughlin for valuable discussions.}
%\begin{thebibliography}{14}

%%%%%%%%%%%%%%%%%%%%%%%%%%%%%%%%%%%%%%%%%%%%%%%%%%%%%%%%%%%
%\balance
% IEEEabrv,
\bibliographystyle{IEEEtran}
%\bibliography{IEEEabrv,Bibliography_X-ray}
\bibliography{Bibliography_CoviLearn}

\section*{About The Authors}

\parpic{\includegraphics[width=1in,clip,keepaspectratio]{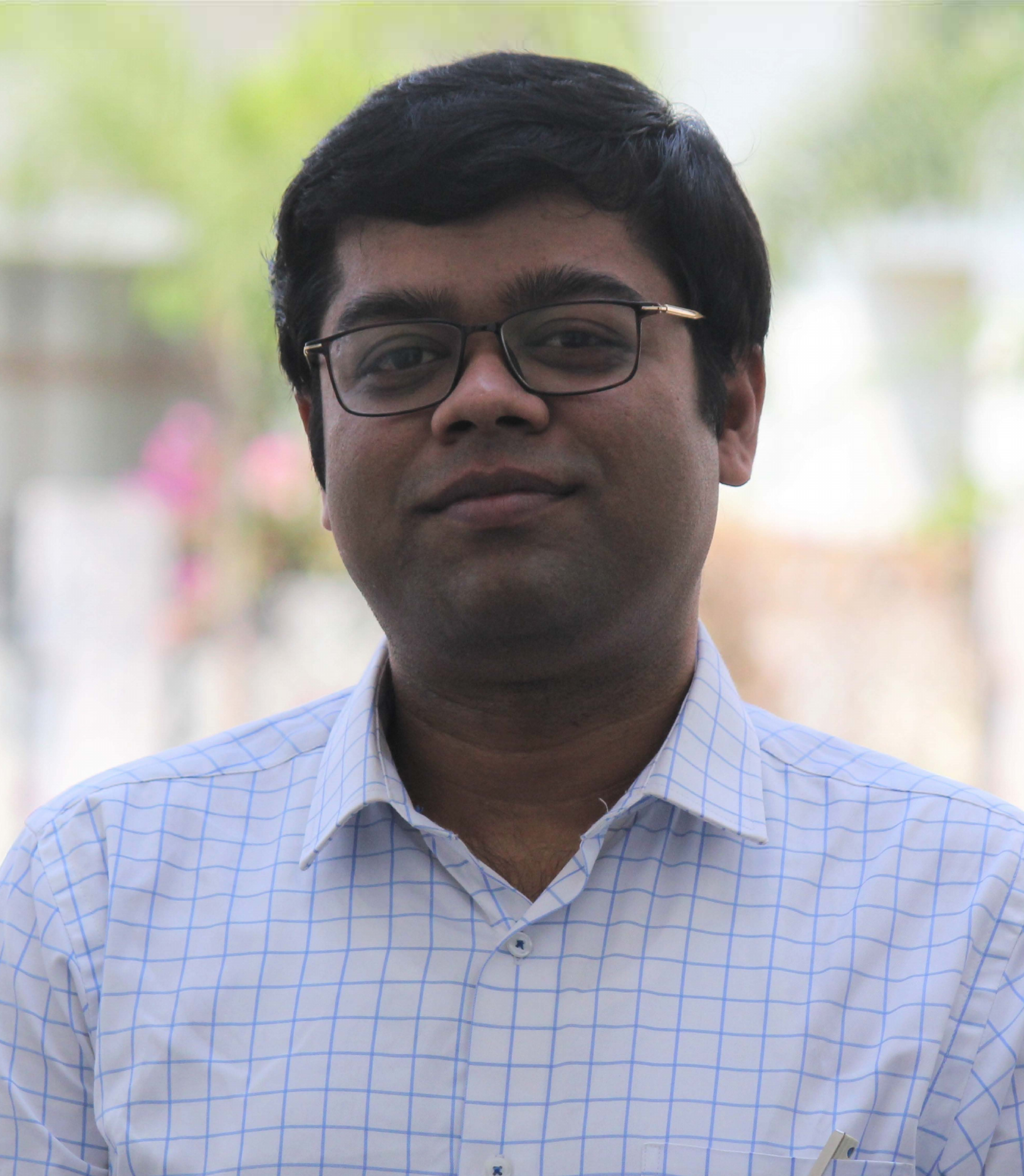}}
\noindent {\bf Debanjan Das} (Senior Member, IEEE) received the B.Tech. degree in applied Electronics and Instrumentation engineering from the Heritage Institute of Technology, Kolkata, in 2009, the M.Tech. degree in Instrumentation from Indian Institute of Technology, Kharagpur, in 2011, and the Ph.D. degree in Electrical Engineering from Indian Institute of Technology, Kharagpur, in 2016. He is an Assistant Professor with Dr. SPM IIIT Naya Raipur. His current research interests include IoT-Smart Sensing, Signal Processing, Bioimpedance, Instrumentation. He has been a member of the IEEE Engineering in Medicine and Biology Society, Measurement and Instrumentation Society.

\par
\vspace{0.5cm}

%\subsection*{ }
\parpic{\includegraphics[width=1in,clip,keepaspectratio]{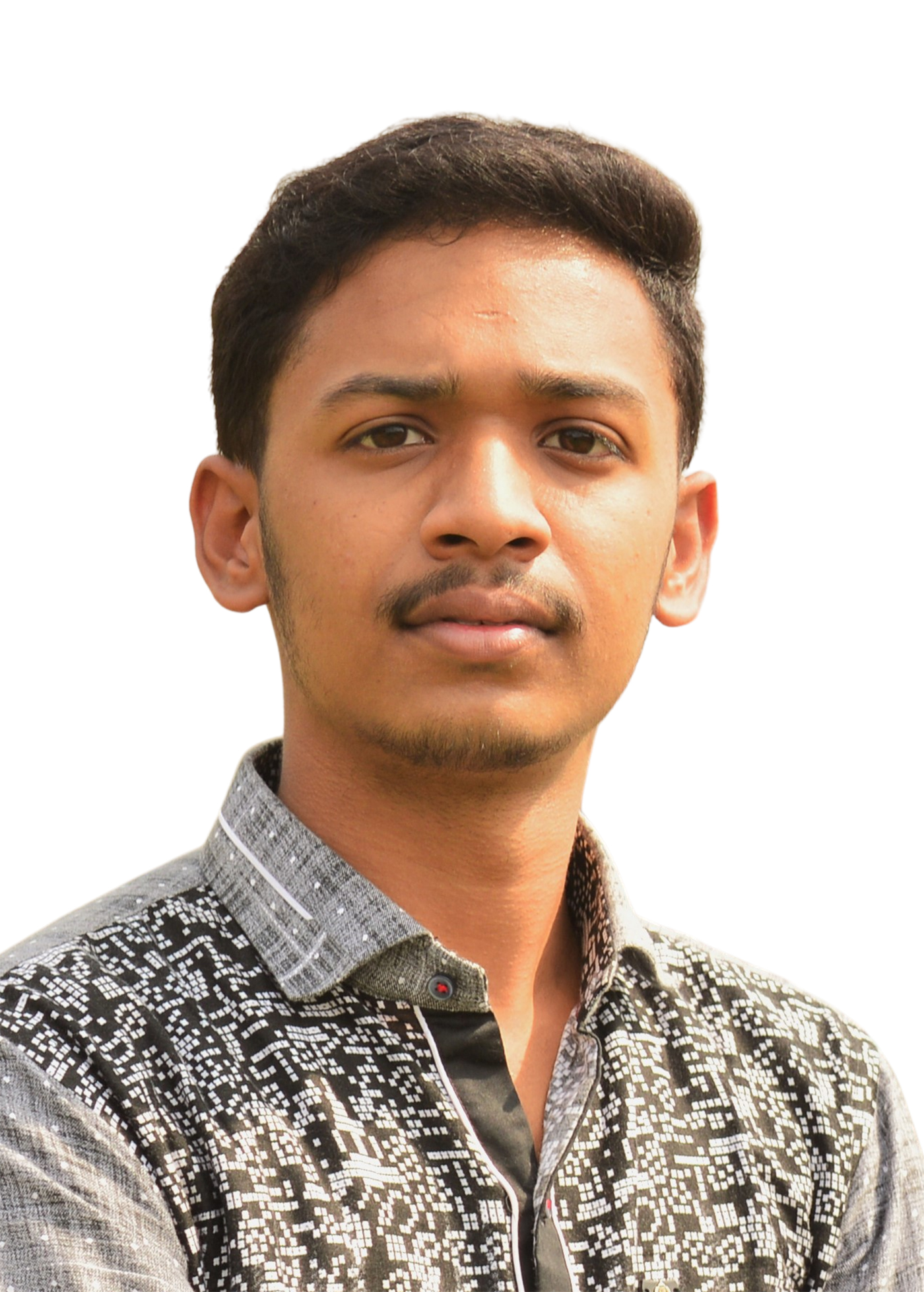}}
\noindent {\bf Chirag Samal}  is an undergraduate student of Electronics and Communication Engineering at the International Institute of Naya Raipur and will be graduating in 2022 with a B.Tech degree in Electronics and Communication Engineering. His research interest  includes Machine Learning, Deep Learning, Pattern Recognition, and Image Processing.
\par
\vspace{2.5cm}

%\subsection*{ }
%\subsection*{ }
%\subsection*{ }
\parpic{\includegraphics[width=1in,clip,keepaspectratio]{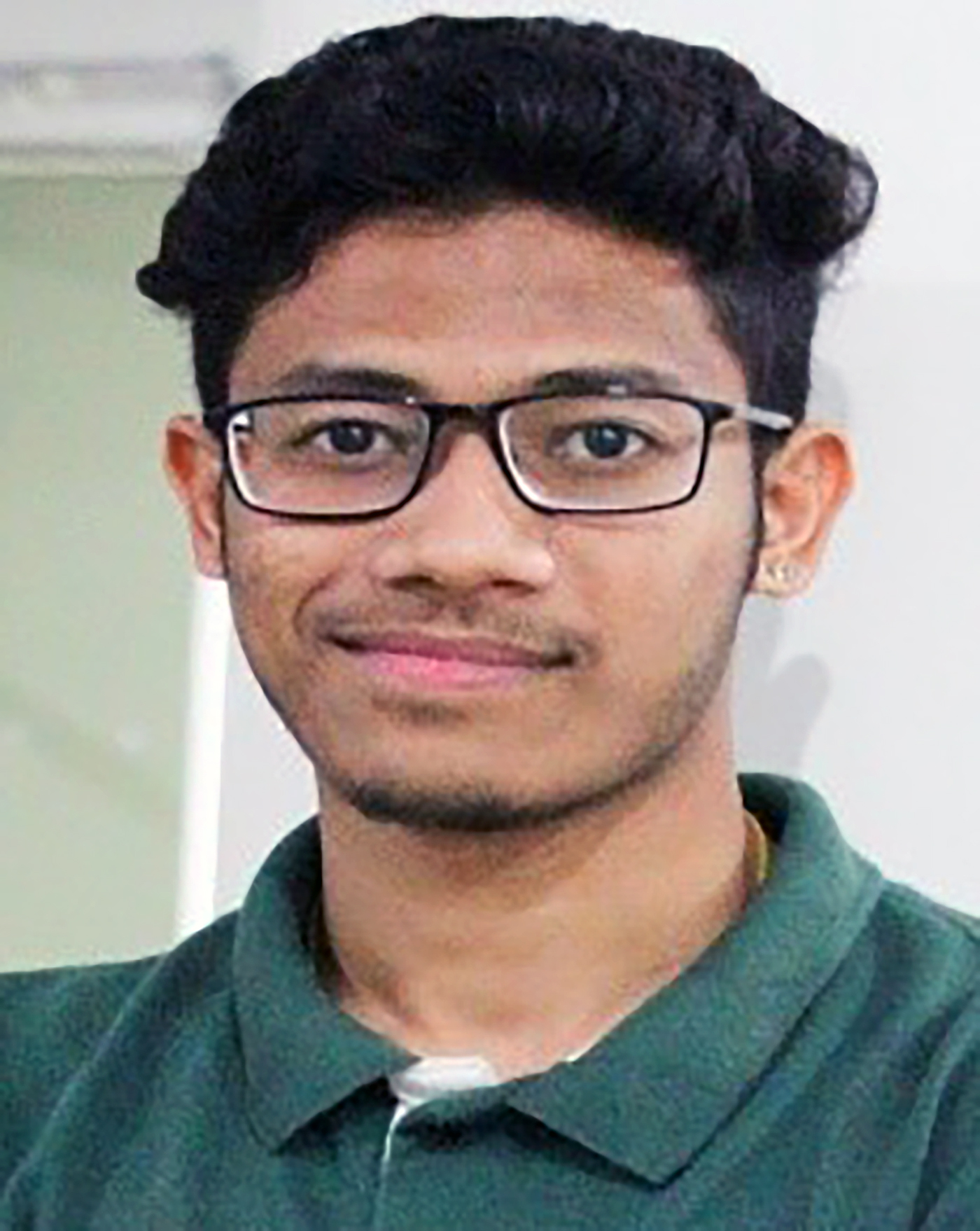}}
\noindent {\bf Deewanshu Ukey} is an undergraduate student of Computer Science Engineering at the International Institute of Information and technology, Naya Raipur and will be graduating in 2022 with a B.Tech degree in Computer Science Engineering. His research interest includes Deep Learning, IoT and AWS Cloud Computing.
\par
\vspace{2.5cm}

%\subsection*{ }
%\subsection*{ }
\parpic{\includegraphics[width=1in,clip,keepaspectratio]{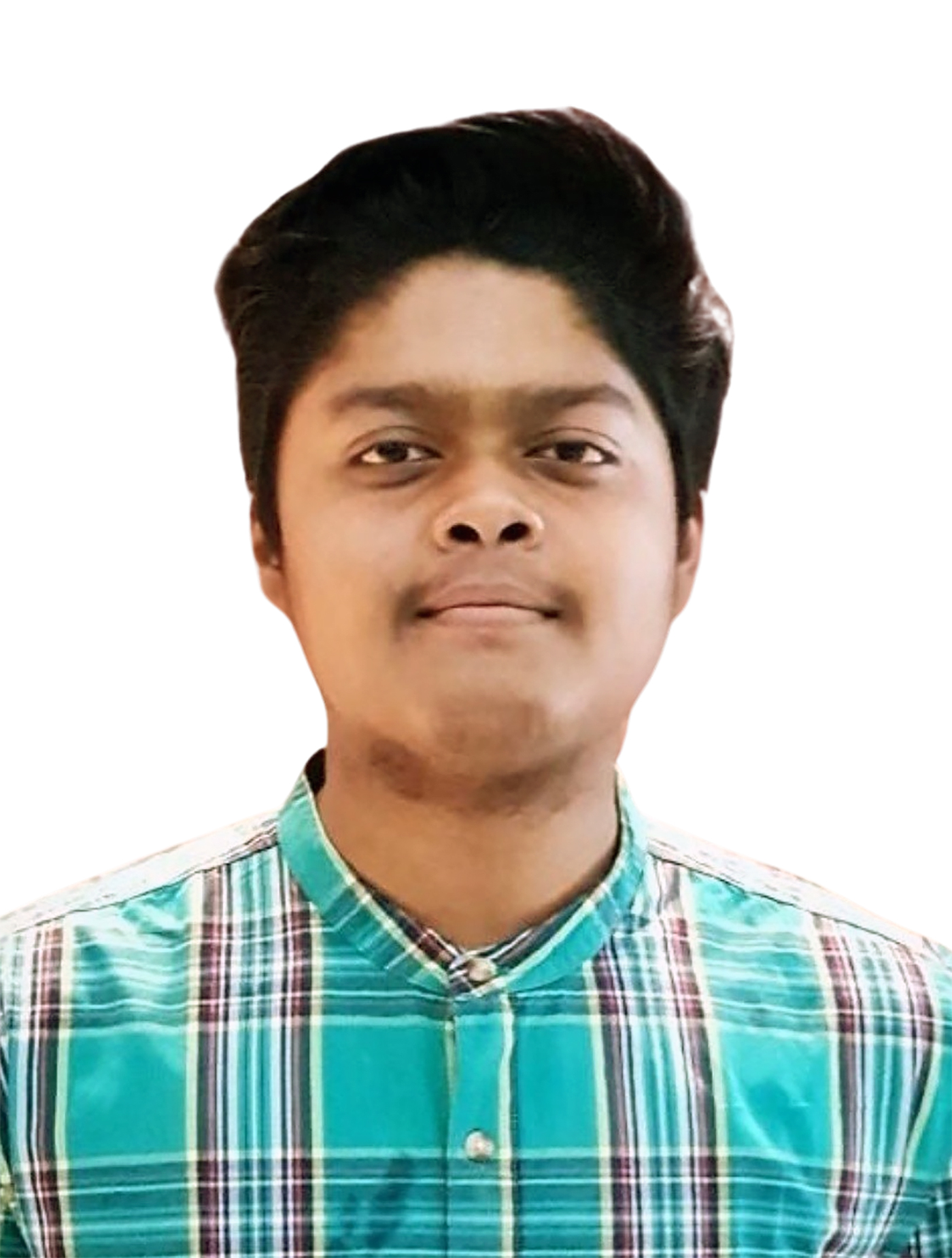}}
\noindent {\bf Gourav Chowdhary} is an undergraduate student of Electronics and Communication Engineering at the International Institute of Naya Raipur and will be graduating in 2022 with a B.Tech degree in Electronics and Communication Engineering. His research interest  includes Machine Learning, Deep Learning, Pattern Recognition, and Image Processing.
\par
\vspace{2.5cm}

%\subsection*{ }
%\subsection*{ }
\parpic{\includegraphics[width=1in,clip,keepaspectratio]{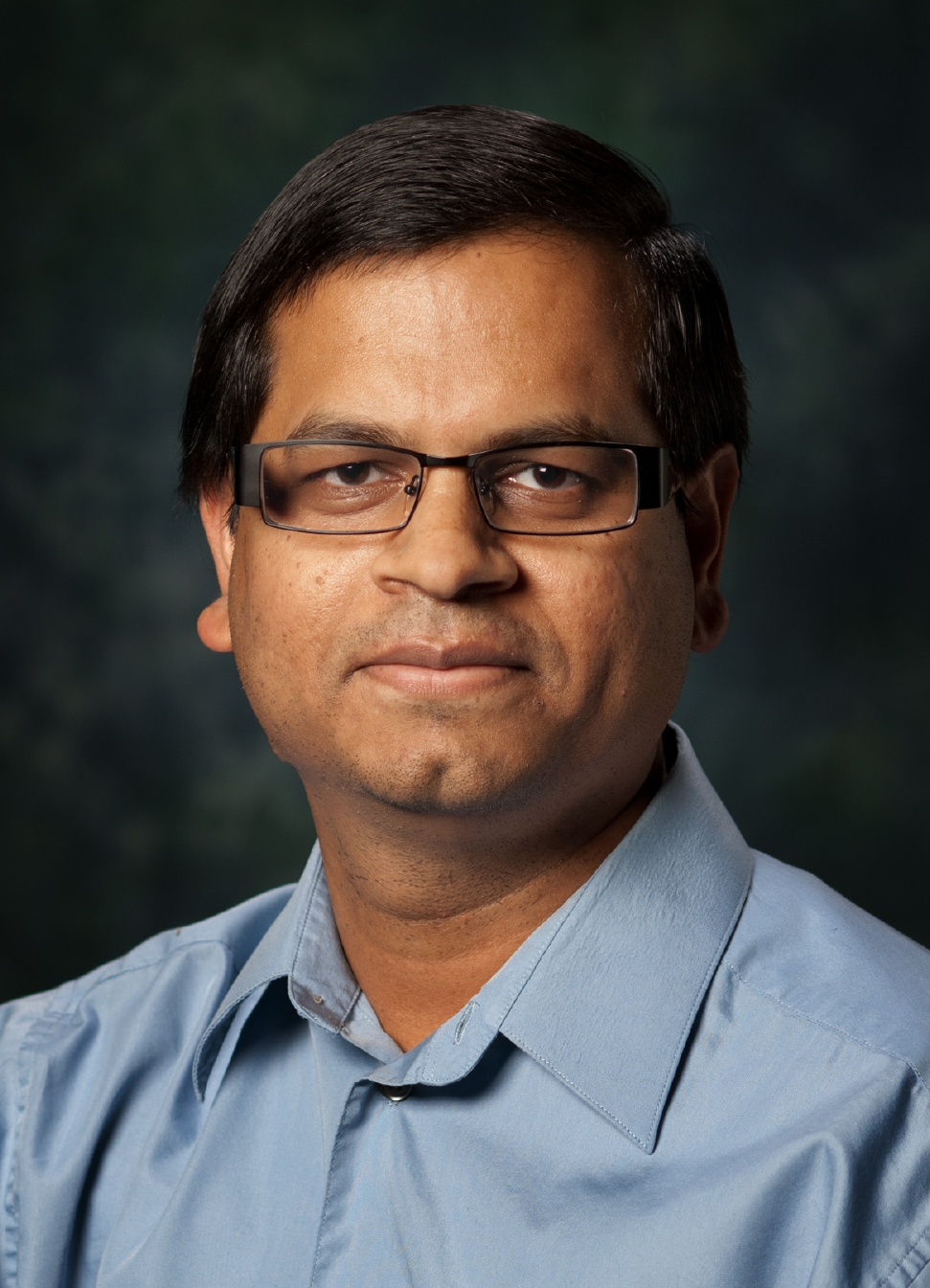}}
\noindent 
{\bf Saraju P. Mohanty} (Senior Member, IEEE) received the bachelor’s degree (Honors) in electrical engineering from the Orissa University of Agriculture and Technology, Bhubaneswar, in 1995, the master’s degree in Systems Science and Automation from the Indian Institute of Science, Bengaluru, in 1999, and the Ph.D. degree in Computer Science and Engineering from the University of South Florida, Tampa, in 2003. He is a Professor with the University of North Texas. His research is in ``Smart Electronic Systems'' which has been funded by National Science Foundations (NSF), Semiconductor Research Corporation (SRC), U.S. Air Force, IUSSTF, and Mission Innovation. He has authored 350 research articles, 4 books, and 7 granted and pending patents. His Google Scholar h-index is 42 and i10-index is 156 with 7300 citations. He is regarded as a visionary researcher on Smart Cities technology in which his research deals with security and energy aware, and AI/ML-integrated smart components. He introduced the Secure Digital Camera (SDC) in 2004 with built-in security features designed using Hardware-Assisted Security (HAS) or Security by Design (SbD) principle. He is widely credited as the designer for the first digital watermarking chip in 2004 and first the low-power digital watermarking chip in 2006. He is a recipient of 13 best paper awards, Fulbright Specialist Award in 2020, IEEE Consumer Technology Society Outstanding Service Award in 2020, the IEEE-CS-TCVLSI Distinguished Leadership Award in 2018, and the PROSE Award for Best Textbook in Physical Sciences and Mathematics category in 2016. He has delivered 11 keynotes and served on 12 panels at various International Conferences. He has been serving on the editorial board of several peer-reviewed international journals, including IEEE Transactions on Consumer Electronics (TCE), and IEEE Transactions on Big Data (TBD). He is the Editor-in-Chief (EiC) of the IEEE Consumer Electronics Magazine (MCE). He has been serving on the Board of Governors (BoG) of the IEEE Consumer Technology Society, and has served as the Chair of Technical Committee on Very Large Scale Integration (TCVLSI), IEEE Computer Society (IEEE-CS) during 2014-2018. He is the founding steering committee chair for the IEEE International Symposium on Smart Electronic Systems (iSES), steering committee vice-chair of the IEEE-CS Symposium on VLSI (ISVLSI), and steering committee vice-chair of the OITS International Conference on Information Technology (ICIT). He has mentored 2 post-doctoral researchers, and supervised 13 Ph.D. dissertations, 26 M.S. theses, and 11 undergraduate projects.

\end{document}